\documentstyle[prl,aps,epsf]{revtex}
\input psfig.tex

\newcommand{\mfig}[4]{ \psfig{figure=#1,width=#2,height=#3,angle=#4 } }

\newcommand{\nota}[1]{ \centerline{ 
                \parbox{14cm}{ { \caption{ {  \footnotesize \sf #1 } \/}}}}}

\begin{document}
\draft
\twocolumn[\hsize\textwidth\columnwidth\hsize\csname
@twocolumnfalse\endcsname

\title{Sporadic randomness, Maxwell's Demon and the Poincar\'{e}
recurrence
times}
\author{Gerardo Aquino$^{1}$, Paolo Grigolini$^{1,2,3}$ and Nicola Scafetta
$^{3}$}
\address{$^{1}$Dipartimento di Fisica dell'Universit\`{a} di Pisa and
INFM, \\
Piazza\\
Torricelli 2, 56127 Pisa, Italy }
\address{$^{2}$Istituto di Biofisica CNR, Area della Ricerca di Pisa,
Via Alfieri 1, San Cataldo 56010 Ghezzano-Pisa, Italy }
\address{$^{3}$Center for Nonlinear Science, University of North Texas,\\
P.O. Box 5368, Denton, Texas 76203 }
\date{\today}
\maketitle

\begin{abstract}
In the case of fully chaotic systems the distribution of
the Poincar\'{e} recurrence times
is an exponential whose decay rate is the Kolmogorov-Sinai (KS)
entropy.
We address the discussion of the same problem, the connection
between dynamics and thermodynamics, in the case of sporadic
randomness, using the Manneville map as a prototype of this class of
processes. We explore the possibility of relating
the distribution of Poincar\'{e} recurrence times
to ``thermodynamics'', in the sense of the KS entropy, also
  in the case of an inverse power law.
This is the dynamic property that Zaslavsly
( G.M. Zaslavsky, Physics Today {\bf 52}(8), 39 (1999)) finds to be
responsible for a striking deviation from ordinary statistical
mechanics under the form of Maxwell's Demon effect.

We show that this way of establishing a connection between
thermodynamics and dynamics is valid only in the case of strong
chaos, where both the sensitivity to initial conditions and the
distribution of the Poincar\'{e} recurrence times are exponential. In
the case of sporadic randomness, resulting at long times
in the L\'{e}vy diffusion processes, the sensitivity to initial
conditions is initially a power law, but it becomes exponential again
in the long-time scale,
whereas the
distribution of Poincar\'{e} recurrence times keeps, or gets, its inverse
power law nature forever, including  the long-time scale where the
sensitivity to initial condition becomes exponential.

We show that a non-extensive
version of thermodynamics would imply
the Maxwell's Demon effect to be determined
by memory, and thus  to be temporary, in conflict with
the dynamic approach to L\'{e}vy statistics. The adoption of
heuristic arguments indicate that this effect is possible, as a form
of genuine equilibrium, after
completion of the process of memory erasure.

\end{abstract}

]

\section{introduction}
In a lucid discussion Lebowitz\cite{lebowitz} has recently restated
the point of view of Boltzmann to establish the microscopic origin of
irreversible macroscopic behavior.
In his view the adoption of the laws of big numbers is essential, and
the role of deterministic chaos becomes important only if it applies
to a macroscopic  number of non-interacting particles.
According to Lebowitz mixing and ergodicity are notions
``unnecessary, misguided and misleading''. In other words, this opinion
reflects the conviction, mirrored by the handbooks  of statistical
mechanics, that
the unification of mechanics and
thermodynamics rests on  $N\rightarrow \infty$, where
$N$ denotes the number of degrees of freedom of the system under
study.

This point of view has to be contrasted with
a new view about the origin of thermodynamics\cite{beck}
produced by the increasing interest in nonlinear dynamics.  Within
this new perspective the
sensitivity to the initial conditions, rather than the large number
of degrees of freedom involved, is playing the crucial role of
generator of thermodynamics.
Zaslavsky\cite{zaslavsky} has recently discussed the phenomenon of the
Poincar\'{e} recurrences without using the traditional argument
  that these recurrences occur
in a so extended time scale as to lie  beyond the observation range.
  Using the Bernouilli shift map, as a dynamic prototype of strong
  chaos, Zaslawsky proves\cite{zaslavsky} that the distribution of
  Poincar\'{e} recurrence times, $P_{R}(t)$, obeys the following
  prescription:
  \begin{equation}
  P_{R}(t) \propto exp(-h_{KS}t),
  \label{exponential}
  \end{equation}
  where $h_{KS}$ denotes the Kolmogorov-Sinai (KS) entropy\cite{beck}.
  Zaslavsky makes the interesting remark that Eq.(\ref{exponential})
  establishes a nice connection between mechanics, the left hand side of this
  equation, and thermodynamics, the right hand side of the same equation. This
  is so because $h_{KS}$ has an entropic significance\cite{beck}.
  It is also well known that the KS entropy is related to the
  Lyapunov coefficient and thus to the sensitivity to initial
  conditions, as a consequence of the Pesin theorem\cite{pesin}.
  To make the connection between $P_{R}(t)$ and the Lyapunov
  coefficient perspective more transparent, it is convenient to
  introduce also the property:
  \begin{equation}
  \xi(t) \equiv \lim_{\Delta x_{0} \rightarrow 0} |\frac{\Delta
  x(t)}{\Delta x(0)}|,
  \label{definition}
  \end{equation}
  where $\Delta x(t)$ denotes the difference between the space
  coordinates of two
  trajectories, moving from two distinct initial conditions at a
  distance $|\Delta x(0)|$ the one from the other. It is evident that
  after identifying $h_{KS}$ with the Lyapunov coefficient $\lambda$,
  we can rewrite Eq.(\ref{exponential}) in the form
  \begin{equation}
  P_{R}(t) \propto \xi(-t) = \frac {1}{\xi(t)} ,
  \label{ambiguity}
  \end{equation}
  where
  \begin{equation}
  \xi(t) = exp(\lambda t).
  \label{exponentialdeparture}
  \end{equation}

  At a first sight, this point of view seems to conflict with that of
  Lebowitz. In this paper we shall argue, with a heuristic approach, that it
is not so,
  and that a sort of compromise between the two views
  can be found. Here we want to limit ourseleves to noticing that
  the observation frequently
  made that the occurrence of strong chaos is exceptional, does not
  necessarily mean that thermodynamics can only rest
  on the ($N\rightarrow \infty$)-perspective. It
  might only imply that the process of transition
  to equilibrium and the ensuing statistics are not
  ordinary. The same aspect might emerge also from within the adoption
  of the ($N\rightarrow \infty$)-perspective. We have two examples in
  mind,
  where the thermodynamic level, if it exists, is closely related to
  anomalous rather than ordinary statistical mechanics.
  The first is the dynamical process recently discussed by Latora
  \emph{et al.}\cite{ruffo}. This is a Hamiltonian system of $N$
  rotors, each rotor being coupled to all the other rotors of the
  system with the same coupling. The authors show that in the limit
  $N\rightarrow \infty$
  from this Hamiltonian picture a process of superdiffusion
  under the form of a L\'{e}vy walk emerges.
  On the other hand, we have in mind a second example, of Hamiltonian
  nature, resulting in L\'{e}vy diffusion. This example is afforded by
  the class of billiards discussed in the recent book of
  Zaslavsky\cite{zaslavsky}. These are systems with only one degree of
  freedom, thereby seemingly departing from the condition $N\rightarrow
  \infty$ that, according to Lebowitz\cite{lebowitz}, is essential for
  the emergence of thermodynamics. In this paper we shall try to prove
  that the point of view advocated by Lebowitz\cite{lebowitz}
  is compatible with that
  sustained by Zaslavsky\cite{zaslavsky,originalpaper,edelman}
  in the case of fully chaotic systems, where
  the dynamical approach to statistical mechanics yields the same ordinary
  form of statistical mechanics as the $(N\rightarrow
  \infty)$-perspective, thereby making apparently useless the adoption
  of a dynamic view. However, we shall argue that the role of dynamics
  might become important when the system under study is characterized
  by persistent memory. We make also the conjecture that a kind of
  statistical equivalence exists between extended memory and
  long-range interactions.

For the time being, we would like to stress
that in his very interesting book\cite{zaslavsky}, as well as in the original
papers\cite{originalpaper,edelman}, Zaslavsky focuses
his attention on the anomalous thermodynamic nature
of these generators of anomalous diffusion. He shows that two
chaotic billiards, coupled to one another through a small hole in the
wall separating one billiard from the other, result in a so strong
violation of
the condition of equal distribution as to suggest
the occurrence of a Maxwell's Demon effect. Maxwell's Demon
was described by Maxwell in his book "Theory of Heat" (1871)
and represented a rather simple device.
Two chambers, $A$ and $B$, are separated by a division with a small hole.
Positioned at the hole the demon allows swifter molecules to pass
from chamber $A$ to $B$ and slower molecules to pass from $B$ to $A$. After
a while chamber $A$ will contain mainly swifter molecules, whereas
chamber $B$ will contain mainly slower molecules.
This implies the appearance of temperature
difference without expenditure of work and contradicts the Second
Law of Thermodynamics.
The version of Maxwell's Demon realized in practice in
Refs.\cite{zaslavsky,originalpaper,today,edelman} is
equivalent to the case where particles with the same kinetic energy
tends to spend more time in one of the two chambers even if the
volume available is the same.

According to Zaslavsky
this surprising behavior is provoked by the fact that the
distribution density $P_{R}(t)$ deviates from the thermodynamic
condition of Eq.(\ref{exponential}). This is so because even in the
phase space of seemingly chaotic systems there might exist stable
islands. The surface of separation between the chaotic sea and the
stable islands is characterized by fractal properties and these
properties make these surfaces very sticky. Consequently, in the
long-time limit the function
$P_{R}(t)$ becomes proportional to $\psi(t)$, denoting the
distribution of waiting times at the border between chaotic sea and
stability island. Using renormalization
group arguments it is shown\cite{zaslavsky} that
\begin{equation}
  \lim_{t \rightarrow \infty} \psi(t) = \frac{const}{t^{\mu}},
  \label{asymptotic}
  \end{equation}
  where $\mu > 2$, in accordance with the Kac theorem\cite{KAC}.
  This means that the first moment of the waiting time distribution is
  finite. After revealing the sources of deviations from the ordinary
  prescriptions of thermodynamics and  statistical mechanics, Zaslasky
  does not rule out the possibility that even in this case a thermodynamic
  perspective applies. He states\cite{today} that all this is obliging
  us to rethink the foundation of thermodynamics.

We see that there are several
  issues to discuss:

  (i) Is there a connection between the condition of strong chaos
  leading to thermodynamics in the
  sense of Eq. (\ref{exponential}) and the Boltzmann
  condition $N\rightarrow \infty$? Is there, in general, a connection
  between the perspective based on deterministic chaos of
  low-dimensional systems and the perspective based on $N \rightarrow \infty$?

  (ii) As far as the connection between  $\psi(t)$ and the sensitivity
  to initial condition is concerned, we see that in the case
  where the exponential picture does not apply
  Eq.(\ref{ambiguity}) splits into the following two
  different possibilities:
  \begin{equation}
  \psi(t) \propto \xi(-t)
  \label{possibility1}
  \end{equation}
  and
  \begin{equation}
  \psi(t) \propto \frac{1}{\xi(t)}.
  \label{possibility2}
  \end{equation}
Which is the correct property?

(iii) Is there a form of thermodynamics involved in this case? If it
is so, which is the nature of this form of thermodynamics? This would
answer the fundamental question raised by Zaslavsky about the origin
of his Maxwell's Demon effect\cite{zaslavsky,today,originalpaper}.

(iv) Is it possible to find a general analytical expression for
$\psi(t)$, fitting the inverse power law of Eq.(\ref{asymptotic}), and
collapsing, for a critical value of a given parameter, into the
exponential form of Eq. (\ref{exponential}) with the same coefficient
$h_{KS}$?

To give the reader a preliminary understanding of question (iv) we
have to make some remarks. We shall
show that the Manneville map\cite{manneville} is a
dynamic process general enough as to reproduce both the exponential
property of Eq.(\ref{exponential}) and the inverse power law of
Eq.(\ref{asymptotic}), depending on the value of a given control parameter
$z$. We shall study this dynamic system while keeping in mind
the
non-extensive thermodynamics introduced by Tsallis in
  1988\cite{CONSTANTINO88}, which is now becoming more and more
  popular\cite{brazil}. From a formal point of view, within this new
  form of statistical mechanics a crucial role is played by the so
  called $q$-exponential. If we express the function $\psi(t)$
  as a $q$-exponential we obtain
\begin{equation}
  \psi(t) = [1 - (1 - q) h_{q} t ]^{1/(1-q)}.
  \label{tsallisproposal}
  \end{equation}
  The parameter $q$ is called entropic index. We see immediately
  that with $q>1$ the function $\psi(t)$ is an inverse power law.
  At the critical value $q =1$ the function
  $\psi(t)$ would make an abrupt transition from the inverse power law
  Eq.(\ref{asymptotic}) to the exponential form of  Eq. (\ref{exponential}).
  An important motivation for this paper has been to assess whether or
  not the non-extensive thermodynamics of Tsallis might be a
  satisfactory solution of question (iii) as well as of (iv).

  This paper is organized as follows. In Section II we discuss the
  dynamical properties of the Manneville map. We shall discuss
  theoretically and numerically the
  waiting time distribution $\psi(t)$ in this specific case.
  In Section III we illustrate very simple arguments supporting the
  choice of Eq.(\ref{possibility2}). Section IV is devoted to exploring
  the possibility
  that the Tsallis non-extensive thermodynamics might shed light on
  the thermodynamic nature
  of the Maxwell's Demon effect. Section V shows that the extended
  regime of transition from dynamics to thermodynamics
corresponds to the steady process of memory erasure
accompanying the dynamic realization of L\'{e}vy processes. Finally
Section VI is devoted to
  making a balance on the answers to questions (i)-(iv) that we are
  giving throughout this paper.

  \section{The Manneville map}
  In this section we illustrate some key dynamic properties
  of the Mannneville map\cite{manneville}. The reader can find surprising
  that after mentioning the key issues of Section I, we
  decide to focus our attention on a non-Hamiltonian system,  with only
  one variable. We feel therefore the need of justifying our choice.
  First of all, we want to remark that, as proved by the authors
  of Ref.\cite{ruffo}, the use of a Hamiltonian system with infinitely
  many degrees of freedom does not rule out the possibility
  that the resulting dynamics, as far as a few collective variables
  are concerned, is equivalent to a random walk process that
  can be obtained dynamically also from a map\cite{geisel}.
  For the emergence of low-dimensional chaos from a Hamiltonian
  picture with a large number of degrees of freedom we refer to the
  discussion of Tennyson \emph{et al.}\cite{tennyson}.

  Then there is another reason why the choice of the Manneville map
  is convenient to mimic the Hamiltonian
  properties responsible for the birth of the
  L\'{e}vy diffusion. According to Zaslavsky, the L\'{e}vy superdiffusion
  in one of his two-dimensional systems is generated by the fact
  that the states of uniform motion last with a waiting time
  distribution that has an inverse power law nature. This means that
  any condition of uniform motion corresponds to a stable island whose
  border with the surrounding chaotic sea is fractal. A trajectory 
sticks to this
  border with an inverse power law distribution of waiting times. For
  the whole sojourn time of the trajectory on this border, in the
  laboratory frame of reference the system undergoes a uniform motion
  in a given direction. From time to time the trajectory leaves the
  fractal region and after a short sojourn in the chaotic sea it sticks
  again to another fractal region. The particle
  sticking to a given fractal region moves in the laboratory frame of
  reference with constant velocity. The change from one sticking
  condition to another might have the effect of changing the motion
  direction, but it does not affect the velocity modulus. Note that the velocity
  modulus is constant because the trajectory motion occurs in the
  constant energy surface of the phase space.
  A nice dynamical model sharing all these properties
  is the egg-carton model of Geisel
  \emph{et al.}\cite{eggcarton}, where the trajectory
  moves of uniform motion along a given channel of the egg-crate
  landscape and at a given time, randomly selected from an inverse power
  law distribution, begins moving in one of the two opposite directions of
  a orthogonal channel. Also this direction of motion is randomly selected.
Notice that \emph{randomly} here refers to a property of
deterministic chaos.

  The one-dimensional version of this
  generator of anomalous diffusion corresponds to a velocity $\xi$
  with only two possible values, $W$ and $-W$. These two states must
  have a distribution of sojourn times with an inverse power law. 
Thus, we see   immediately that the one-dimensional
  version of the Hamiltonian systems studied by Zaslavsky becomes
  similar to the
  one-dimensional map of Ref.\cite{geisel}. This map has been widely used in
  the recent past to derive L\'{e}vy
  processes\cite{zumofen1,zumofen2,trefan,allegro}.  On the other hand,
  for the purpose of the discussion of the present paper, as already
  done in the earlier work of Ref.\cite{ANNA}, it is convenient to
  shift our attention from the bimodal maps of
  Refs.\cite{geisel,zumofen1,zumofen2,trefan,allegro} to the case of
  the Manneville map, which has only one laminar region.
  The Kolmogorov complexity of the Manneville map has been already
  analyzed years ago by Gaspard and Wang\cite{gaspard}. The complexity
  of a sequence equivalent to that of the bimodal maps of
  Refs.\cite{geisel,zumofen1,zumofen2,trefan,allegro} was studied
  recently by Buiatti \emph{et al.}\cite{marco}. It is straightforward
  to show that the Kolmogorov complexity of the Manneville map is
proportional to
  the complexity of the bimodal maps generating L\'{e}vy
  processes\cite{note}.
We are interested in sporadic randomness, and for
  our purposes it is irrelevant whether after exiting from one laminar
  region and crossing the chaotic region, the trajectory is injected
  back into the original laminar region, or into a new laminar region
   equivalent to the original\cite{note}. If we make the arbitrary
   decision that crossing the chaotic region always implies a
   transition from one laminar region to the other, the resulting
   complexity is the same as that of the Manneville map, with only one
   laminar region, and consequently no uncertainty associated with its
   choice. In the  bimodal case,  the injection back to
   the original laminar region and the jump into the other laminar
   region are random processes,
   with the same probability.  This implies a Kolmogorov complexity
   which is twice as that of the Manneville map\cite{note}.

   In conclusion, the reason for
  the choice of the Manneville is that this map mimics very well the condition
  of sporadic randomness that, according to Zaslavsky\cite{zaslavsky},
  is responsible for Maxwell's Demon effect. Therefore, we are
  convinced that a numerical treatments of the Hamiltonian systems
  with sporadic randomness would lead to the same conclusions as those
  reached by us here, on the basis of the Manneville map.

The Manneville map reads:
\begin{equation}
   x_{n+1} = \Phi(x_{n}) = x_{n} + x_{n}^{z} (mod.1) (1\leq z).
   \label{manneville}
   \end{equation}
   We note that at $z = 1$ the Manneville map becomes equivalent to the
   Bernouilli shift map used by Zaslavsly to prove the fundamental
   result of Eq. (\ref{exponential}). For $z>1$ the interval
   $[0,1]$ is divided in two regions, the laminar region, $[0, d(z)]$,
   and the chaotic region, $[d(z),1]$, with $d(z)$ defined by
   \begin{equation}
   d(z) + d(z)^{z} = 1
   \label{definitionofd}
   \end{equation}

   We review here the arguments used by Geisel and Thomae\cite{geisel}
   to derive an analytical expression for the distribution of the
   times of sojourn in the laminar region. First of all we assume that
   the injection point $x_{0}$ is so close to $x = 0$ as to replace
   eq.(\ref{manneville}) with:
   \begin{equation}
   dx/dt = x^{z}.
   \label{continuouspicture}
   \end{equation}
   Thus we obtain the following time evolution:
   \begin{equation}
   x(t) = [x_{0}^{1-z} + (1-z)t]^{1/(1-z)}.
   \label{timevolution}
   \end{equation}
   Hence the time necessary for the trajectory to get the border $x =
   d(z)$ is given by
   \begin{equation}
   t = T(x_{0}) \equiv (\frac{1}{x_{0}^{z-1}}-\frac{1}{d^{z-1}})\frac{1}{z-1}.
   \label{timetoreachtheborder}
   \end {equation}
   Note that in the special case where the initial conditon is so
   close to $x = 0$ as to fulfill the condition $x_{0}<< d$, the exit
   time $T(x_{0})$ can be satisfactorily approximated by
   \begin{equation}
       T(x_{0}) \approx \frac{1}{x_{0}^{z-1}}\frac{1}{z-1},
       \label{instability}
       \end{equation}
       which, as we shall see in Section III, can be used to define the
       time at which we lose control of the trajectories departing from
       a region of the map very close to $x = 0$.

   Note that the distribution function $\psi(t)$ is related to the
   injection probability $p(x_{0})$ by
   \begin{equation}
   \label{distributiondistribution}
   \psi(t)dt = p(x_{0})dx_{0}.
   \end{equation}
   Assuming equiprobability for the injection process we have:
   \begin{equation}
   p(x_{0})dx_{0} = \frac{1}{d(z)}|\frac{dx_{0}}{dT}| dT.
   \label{equiprobability}
   \end{equation}
   Thus we obtain:
   \begin{equation}
   \psi(t) =  \frac{1}{d(z)}|\frac{dx_{0}}{dt}|.
   \label{connection}
   \end{equation}
   By differentiating $t$ with respect to $x_{0}$ we finally arrive at
   \begin{equation}
   \psi(t) = d^{z-1} [1 + d^{z-1}(z-1)t]^{-z/(z-1)}.
   \label{waitingfunction}
  \end{equation}
  It is important to observe that the mean waiting time $T_{av}$ is
  given by
  \begin{equation}
      T_{av} = \frac{1}{d^{z-1}}\frac{1}{2-z},
      \label{meanwaitingtime}
      \end{equation}
      with $T_{av} = \infty$ for $2 \leq z$

  We see that Eq.(\ref{waitingfunction}) implies that the
  region corresponding to $z >2$ is characterized by a diverging first
  moment. This region is in conflict with the Kac theorem\cite{KAC}
  and for this reason we do not take it into account. The region
  pertaining to the interval $1.5 < z <2$ is characterized by a finite
  second moment and a diverging second moment. This is the region of
  interest for us, since it corresponds to that generating
  L\'{e}vy diffusion according to the recent work of Ref.\cite{ANNA}.
  The region $1<z<1.5$ is characterized by a finite second moment. Of
  course, the ideal condition $z = 1$ implies all the moments to be
  finite. This is a region which would correspond, in the perspective
  of Ref.\cite{ANNA}, to ordinary Gaussian diffusion.
  We note that  in the region $ 1< z< 1.5$ the waiting
  function distribution must make a transition from the inverse power
  law behavior of Eq.(\ref{waitingfunction}) to the exponential regime,
  where the arguments of Zaslavsky leading to Eq.(\ref{exponential})
  apply. In fact, at $z=1$, the Manneville map becomes identical to
  the Bernouilli shift map. In that case, the theoretical remarks of
  Zaslavsky yield:
  \begin{equation}
  \psi(t) \propto exp(- tln2).
  \label{strongchaos}
  \end{equation}

  We are now in a position to make a preliminary observation
  about question (iv) of Section I.
The analytical expression of Eq.(\ref{waitingfunction}) can be
expressed in the form of the $q$-exponential of
Eq.(\ref{tsallisproposal}), with
  \begin{equation}
  q = 1 + (z-1)/z .
  \label{laterdiscussed}
  \end{equation}
  However, this would not fit completely the request of being an exact
  representation of the waiting function in the fully chaotic case. In
  fact $q=1$ would correctly imply $z = 1$, but this would make
  Eq.(\ref{waitingfunction}) collapse into
  \begin{equation}
  \psi(t) \propto exp(- t).
  \label{wrongstrongchaos}
  \end{equation}
  rather than in the exact expression of Eq.(\ref{strongchaos}).

This, in principle,
would not rule out the possibility
of finding a different analytical expression
fulfilling the requirement of question (iv).
To explore more deeply issue (iv) we use a numerical treatment of the 
Manneville
  map of Eq. (\ref{manneville}). For computational purposes we found to
  be more convenient to evaluate the population of the laminar region,
  $M(t)$, rather than the waiting time distribution $\psi(t)$. The two
  functions are related the one to the other by

  \begin{equation}
  M(t) = 1 - \int_{0}^{t} \psi(t')dt'.
  \label{fromMtopsi}
  \end{equation}
  Thus, the analytical expression of Eq.(\ref{waitingfunction}) yields
  \begin{equation}
  M(t)  = \left[1 + d^{z-1}(z-1)t\right]^{-1/(z-1)}.
  \label{explicit}
  \end{equation}
  The explicit expression of $M(t)$ in the fully chaotic case (z = 1)
  is derived from Eq.(\ref{exponential}). In fact, Zaslavsky
  proves\cite{zaslavsky} that Eq.(\ref{exponential}) becomes exact in
  the case of the Bernouilli map, with $h_{KS}= ln 2$. In the
  normalized form, Eq.(1) becomes:
  \begin{equation}
      \psi(t) = h_{KS} exp(-h_{KS}t).
      \label{normalized}
      \end{equation}
      Let us plug Eq.(\ref{normalized}) into Eq.(\ref{fromMtopsi}). We
      obtain
  \begin{equation}
  M(t) = exp(-t ln2).
  \label{usual}
  \end{equation}

  We know that Eq.(\ref{usual}) is exact and we expect
  Eq.(\ref{explicit}) to become exact for $z \rightarrow \infty$.
  In conclusion, as to the important function $\psi(t)$, we know its
  exact expression at $z = 1$ and in the asymptotic limit
  $z \rightarrow \infty$. It is
worth exploring the intermediate region with
  a numerical treatment.

  Figs. 1 and 2 illustrate the results of our numerical treatment.

\begin{figure}
\centerline{ \mfig{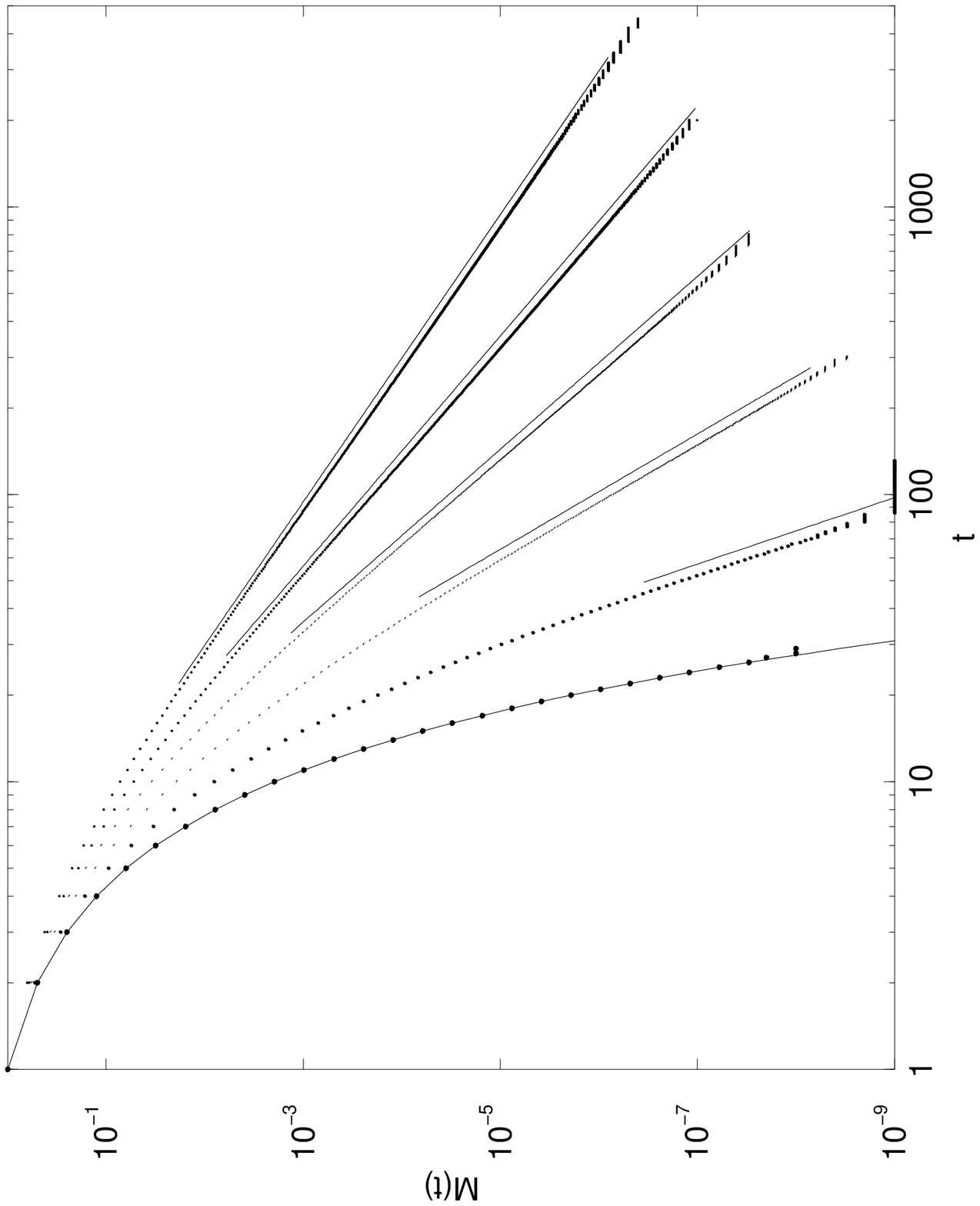}{7.5cm}{6cm}{270}}
\nota{\label{label1}M(t) as a function of time. The meaning of the six full lines is as
  follows. The lowest full line is the function $M(t) = exp(-t ln2)$.
  All the other full lines denote the long-time inverse power law
  $M(t) = 1/t^{\frac{1}{z-1}}$. The dotted lines are the numerical
  result. All the full lines but the lowest have been shifted to the
  right to make them distinguishable from the numerical result.
  The value of the parameter $z$, from
  the bottom to the top is: $z = 1, 1.1, 1.2, 1.3, 1.4, 1.5$}
\end{figure}

\begin{figure}
\centerline{ \mfig{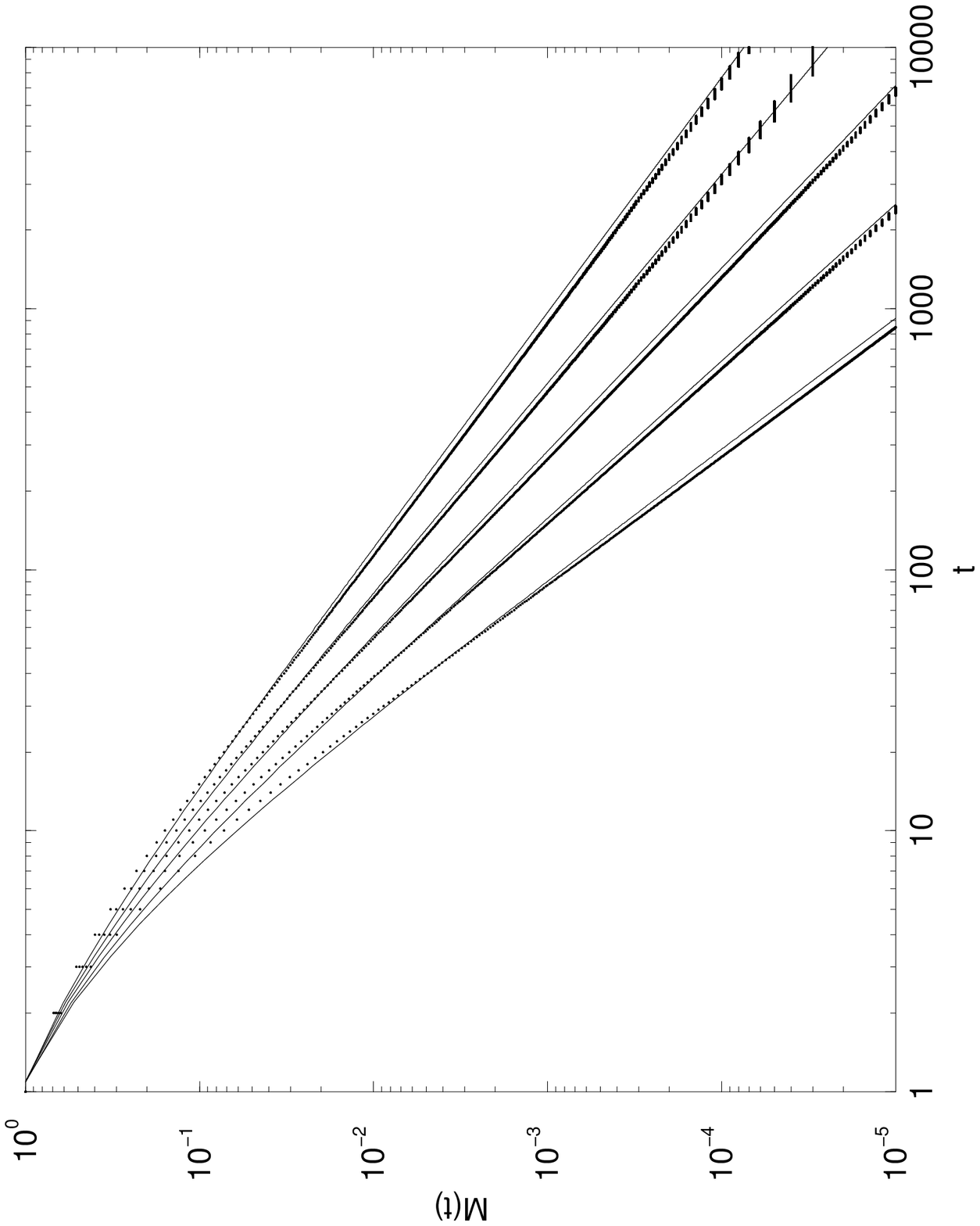}{7cm}{5.5cm}{270}}
\nota{\label{label2} M(t) as a function of time. The meaning of the  four pair of lines is
  as follows. The  full lines denote
  the function $M(t)$ of  Eq.(\ref{explicit}) and the dotted lines 
denote the numerical results.
  To make he full lines distinguishable from the dotted lines we
  shifted them to the right by the quantity
  $\epsilon =0.1$. In the logarithmic representation adopted, this is
  equivalent to replacing $t$ of $M(t)$ with $t exp(-\epsilon)$. The
  value of the parameter $z$ from the bottom to the top changes as
  follows: $z = 1.5, 1.6. 1.7, 1.8,1.9$}
\end{figure}

  In Fig. 1 we illustrate the result of the numerical calculation with
  the parameter $z$ in
  the interval $[1, 1.5]$.
  We see that the long-time limit fits for all  values of $z$ considered
  but $z=1$, the theoretical prescription of the inverse power law
  $t^{-\frac{1}{z-1}}$. However, this inverse power law regime is
  reached after an extended transition regime, which seems to be
  exponential-like. The duration of this transition regime become more
  and more extended with $z$ coming closer and closer to $z=1$. At $z=
  1$ this transition regime becomes infinitely extended and coincident
  with the theoretical prediction of Eq.(\ref{usual}).

  In Fig. 2, devoted to studying the relaxation of the Manneville map
  with $z$ in the interval $[1.5,2]$, we see that the prediction of
  Eq.(\ref{explicit}) is very
  accurate and tends to become exact with increasing the value of $z$.

  In conclusion, these numerical results prove that the $q$-exponential of
  Eq.(\ref{tsallisproposal}) cannot satisfactorily fit the
  conditions set by question (iv).
  However, this is not yet a strong reason to rule out
  the non-extensive thermodynamics of
  Tsallis\cite{CONSTANTINO88,brazil}. Let us make another attempt
  at establishing this non-extensive thermodynamics as
  the proper form of thermodynamics that Zaslavsky is looking for.
  We now set a request more flexible than the one earlier adopted.
  This is based on disregarding
  the extended regime of transition to the inverse power law. In other
  words, we now accept an analytical form of $q$-exponential which at
  $q=1$ becomes identical to Eq.(\ref{strongchaos}) while at $z > 1$
  results in the same inverse power law as the Manneville map.
  We try with the following possibility:
    \begin{equation}
  \psi(t) = [1 + (1 - q) h_{KS}t ]^{1/(1-q)},
  \label{tsallisproposal2}
  \end{equation}
  supplemented by Eq.(\ref{laterdiscussed}).
  To a first sight, this possibility looks very attractive. In fact, at
  $q=1$, the prescription of Eq.(\ref{laterdiscussed}) yields $z=1$.
  This means the Bernouilli map, and, consequently\cite{beck}
  \begin{equation}
  h_{KS} = ln 2.
  \label{standardks}
  \end{equation}
  Thus, the requirement of recovering Eq.(\ref{strongchaos}) is
  fulfilled, while the request of producing the correct power law at
  $z>1$ is obviously satisfied due to the Eq.(\ref{laterdiscussed}).

  However, this is not physically correct.  We let the reader know
that years ago,  Gaspard and Wang\cite{gaspard} evaluated the
  Lyapunov coefficient of the Manneville map
and found it
to be given with a good accuracy by $h_{KS} = (2-z)z ln2$. This means
that $h_{KS}$ vanishes at $z=2$, thereby yielding a strong
disagreement between Eq.(\ref{tsallisproposal2}) and
Eq.(\ref{waitingfunction}).

One might think that the departure of Eq.(\ref{tsallisproposal2}) from
Eq.(\ref{waitingfunction})  is caused by the fact that
the coefficient $h_{q}$ appearing in
Eq.(\ref{tsallisproposal}) cannot be identified with
the ordinary Lyapunov coefficient. In fact, according
to the Pesin theorem\cite{pesin} we have
\begin{equation}
     h_{KS} = \lambda \equiv \int_{0}^{1} dx \rho(x) ln (d\Phi/dx).
     \label{ordinary}
     \end{equation}
     One might think that the adoption of Tsallis non-extensive
     thermodynamics implies that the ordinary definition
     of  Lyapunov
     coefficient cannot be maintained and that consequently $h_{q}$ 
cannot be identified
     with the property of Eq.(\ref{ordinary}), which is in fact the
     property discussed by Gaspard and Wang\cite{gaspard}. In Section
     IV B we shall prove that also this possibility has to be ruled out.
    At the same time we shall show that the correct answer to question
    (ii) cannot be given by Eq.(\ref{possibility1}).
     The derivation of the correct answer to
question (ii) is given in Section III.

  \section{sensitivity to initial conditions}
  The sensitivity to initial conditions of the Manneville map has been
  studied in Ref. \cite{ANNA}. The authors of this paper evaluated the
  function $\xi(t)$ defined by Eq.(\ref{definition}), using the
  same assumptions as those adopted in Section II to derive Eq.
  (\ref{waitingfunction}). The result is given by:
  \begin{equation}
  \xi(t) = [1 -(z-1)x_{0}^{z-1}t]^{-z/(z-1)}.
  \label{superdiffusion}
  \end{equation}
  This function is, in a sense, faster than the exponential, since it
  diverges at the time $T_{div}$ given by
  \begin{equation}
  T_{div} \equiv \frac{1}{(z-1) x_{0}^{z-1}},
  \label{strongerthanexponential}
  \end{equation}
  while the case of exponential sensitivity would yield
  finite values of $\xi(t)$ for any finite value of time.

  It would be tempting to conclude that the Manneville map is
  characterized  by a form of sensitivity to initial condions stronger than the
  usual exponential sensitivity, but it  would be wrong. Let us
  explain why it is so using heuristic arguments. Let us divide
  the phase space of the Manneville map into two regions. The first is
  very close to $x = 0$, and it is characterized
  by the the local Lyapunov coefficients
  \begin{equation}
      \lambda(x_{0}) = ln(1+zx_{0}^{z-1}) \approx zx_{0}^{z-1}.
      \label{firstregion}
      \end{equation}
      The second region includes the whole chaotic region plus a
      portion of the laminar region and it is characterized by
      local Lyapunov coefficients of the order of
      \begin{equation}
	 \lambda \approx ln (1 + z).
	 \label{secondregion}
	 \end{equation}
       Let us now define the
  time at which we lose control of a trajectory, in the case of
  exponential
  sensitivity, as a time proportional to
  $1/\lambda$, where $\lambda$ is the Lyapunov coefficient.
  We see that the time at which we lose control of the trajectories
  departing from the first region is of the order of the time scale
  $T_{div}$, of Eq.(\ref{strongerthanexponential}),
  while the time at which we lose control of the trajectories
  departing from the second region is of the order of unity. We note
  that for values of $x_{0}$ very close to $x=0$ we can fulfill the
  inequality:
  \begin{equation}
   T_{av} << T_{div},
      \label{crucialinequality}
      \end{equation}
      where $T_{av}$ is the waiting mean time defined by
      Eq.(\ref{meanwaitingtime}).
      Note also that for $x_{0}$ so small as to
      fit the condition of Eq.(\ref{crucialinequality}), the time
      $T_{div}$ can be interpreted in two distinct ways. The former
      is the time at which $\xi(t)$ diverges, and the second is the
      inverse of the local Lyapunov coefficient of
      Eq.(\ref{firstregion}).
      We see, in conclusion, that the complexity of the Manneville map
      is decided by the time scales $1/log(1+z)$ and $T_{av}$ rather
      than by $T_{div}$.
  This is the reason why the randomness detected
  by the method of Kolmogorov complexity\cite{gaspard}
  is strongly determined by the influence of the chaotic region of the
  Manneville map.

  This is a key point worth of a comment. The function $\xi(t)$ with
  the form of Eq.(\ref{superdiffusion}) cannot be used for the
  statistical interpretation of the dynamic process under study
  for times larger than the exit times, and consequently
  cannot have any relevance when the time $T_{av}$ of
  Eq.(\ref{meanwaitingtime}) is finite, namely for $z < 2$. In this
  case, in the time scale
  $t >> T_{av}$  the analytical form of the
  function $\xi(t)$ changes and becomes exponential\cite{gaspard}. At
  this stage there are no more possible
  connections between $\psi(t)$ and $\xi(t)$ left.
  In fact,the function $\psi(t)$ has a meaning
  completely different
from that of $\xi(t)$. The function $\psi(t)$ signals the probability
of observing sojourn
  times of any length, including those exceeding the time scale $T_{av}$.
  Thus, the function $\psi(t)$ maintains its inverse power law nature
  even in the time scale where $\xi(t)$ is exponential.

  In accordance with this remarks we  find a connection
  between the function $\psi(t)$ and the function $\xi(t)$ at the exit
  time $T$ using Eq.(\ref{possibility2}).
  The exit time  depends on the initial condition of the
  trajectory within the laminar region and is given by
  Eq. (\ref{timetoreachtheborder}).
  Let us express $x_{0}$ of Eq.(\ref{superdiffusion}) in terms of this
  exit time, and let us replace it into Eq. (\ref{superdiffusion}).
  The resulting expression has to be interpreted as the value that
  the function $\xi(t)$ gets at the exit time. Its explicit
  expression is:
  \begin{equation}
  \xi(T) = d^{z}[d^{1-z} + (z-1)T]^{z/(z-1)}.
  \label{gerardo}
  \end{equation}
  Comparing Eq.(\ref{gerardo}) to Eq.(\ref{waitingfunction}) we reach
  the conclusion that:
  \begin{equation}
  \psi(T) \propto \frac{1}{\xi(T)}.
  \label{inverseprop}
  \end{equation}
  Thus with very simple arguments we are led to
  conclude that the generalization of Eq.(\ref{ambiguity}) implies the
  choice of Eq.(\ref{possibility2}). The meaning of this conclusion is
  that the value of $\psi(T)$ is determined by that of the function
  $\xi(t)$ with $t$ being a given exit time.
  This very simple argument settles question ($ii$) of
  Section 1. We have to address now the much more delicate question of
  the connection with the thermodynamic perspective.
  Before ending this section, we would like to notice that the
  condition of Eq.(\ref{inverseprop}) does not apply only in the region
  $z > 1.5$ which favors the continuous approximation behind the
  analytical results of Eqs.(\ref{superdiffusion}) and
  (\ref{waitingfunction}). It applies very well also to the long-time
  limit in the region
  $1<z<1.5$, as clearly illustrated by Fig. 1.

  \section{thermodynamic approach to the waiting time distribution}

  Here we show that the entropic arguments of the non-extensive
  thermodynamics\cite{brazil} would naturally yield
  Eq.(\ref{possibility1}) rather than Eq.(\ref{possibility2}), creating
  a conflict with the conclusions of Section III. Thus the entropic
  arguments yield a wrong conclusion, which
by itself does not necessarily imply that the
  Tsallis non-extensive thermodynamics is ill founded. If the Tsallis
  non-extensive thermodynamics  were well
  founded, on the other hand, this wrong conclusion would certainly
  imply that this form of thermodynamics\cite{CONSTANTINO88,brazil} is 
not permanent.
  In Section V we shall show, on the other hand, that the Maxwell's
  Demon effect is permanent and takes place in a later time scale .

  \subsection{Entropic derivation of $\psi(t)$}
  The dynamic derivation of L\'{e}vy processes from intermittent 
  maps\cite{trefan,allegro}
  has established that probability for the particle to make a jump of
  length $|x|$ in a time t, $\Pi(x)$, is related to $\psi(t)$ by
  \begin{equation}
      \Pi(x) = \psi(x/W)/W.
      \label{pi2}
      \end{equation}
      To make immediately evident to the reader the explanation of the
      form of
      Eq.(\ref{pi2}) it is enough for us to let him/her know that the dynamic
      derivation of of L\'{e}vy processes\cite{trefan,allegro} rests on
      a one-dimension motion with a fluctuating velocity. The
      particle velocity fluctuates among the values $W$ and $-W$. The
      particle sojourns in these two states with a distribution of
      waiting times given by $\psi(t)$, which has the same structure
      as that of Eq.(\ref{waitingfunction}). This accounts for
      Eq.(\ref{pi2}). On the other hand, on the basis of the increasing
      interest for the subject of non-extensive entropy\cite{brazil},
      and along lines similar to those adopted by earlier
      work\cite{AZ94,ZA95,TLSM9}, we decide to maximize the Tsallis
      entropy\cite{CONSTANTINO88} to determine the form of $\Pi(x)$.
Therefore we define

  \begin{equation}
\label{tsentr}
S_{q}[\Pi(x)] = - \frac{ 1- \int_{-\infty}^{\infty}{\Pi(x)}^{q}dx}{(1-q)} ,
\end{equation}
and we maximize it under the constraint of assigning a fixed value to
the first moment of the escort distribution with entropic index $q$.
This means that, as done in Ref.\cite{ANNA}, we have to
maximize the functional form $F_{q}(\Pi)$ defined by

\begin{equation}
\label{fun}
F_{q}(\Pi)\equiv  \frac{ 1- \int_{-\infty}^{\infty}{\Pi(x)}^{q}dx}{(1-q)}
- \beta \int_{-\infty}^{\infty}|x| {\Pi(x)}^{q} dx + \alpha
\int_{-\infty}^{\infty} \Pi(x) dx .
\end{equation}
As a result of the entropy maximization we get\cite{ANNA}
\begin{equation}
\label{TRANSITION}
\Pi(x)={A\over{[1+\beta(q-1)|x|^{\nu}]^{1/(q-1)}}}.
\end{equation}

It is evident that for this  entropic argument to be compatible
with the statistical derivation of $\psi(t)$ as resulting from
Eq.(\ref{waitingfunction}) we are forced to accept the
relation of Eq.(\ref{laterdiscussed}).
In the region $1 \leq z \leq 1.5$, as shown in Fig. 1, the time regime
of validity of the inverse power law fulfilling Eq.(\ref{laterdiscussed})
is confined to regions of larger and larger times as $z \rightarrow 
1$, since the
size of the early time region with exponential-like behavior becomes more
and more extended.

We notice that the Tsallis information approach affords a formal
justification for the birth of an inverse power law. However,
the special value of Eq.(\ref{laterdiscussed}) is apparently established in
such a way as to make the entropic argument compatible with the
dynamic argument, with \emph{ad hoc} arguments, rather
than as a result of a theoretical prediction. The role of the
entropy of Eq.(\ref{tsentr}) would become much more important if it
were possible to use it with no recourse to \emph{ad hoc} arguments.
In Section IV B we shall show that perhaps it
is possible to establish a dynamic approach to this
non-extensive thermodynamics, and consequently to a $q >1$ established
with no \emph{ad hoc} assumption,
with the caution, however, of interpreting this thermodynamics as 
temporary. We shall discuss also in which sense  the entropic index
given by Eq.(\ref{laterdiscussed}) as to be thought of as ``magic''.

\subsection{A transient form of Kolmogorov-Sinai entropy}
The results found years ago by Gaspard and Wang\cite{gaspard}
that the ordinary Lyapunov coefficient is finite in the range
$1\leq z \leq 2$ seem to rule out the possibility of
applying to the present case the prediction of Eq.(\ref{exponential}).
In fact, since $h_{KS}$ is finite, we can use Eq.(\ref{exponential}),
but this would lead to an exponential $\psi(t)$ in deep contrast with
the fact that is an inverse power law instead.
The breakdown of this important theoretical prediction is certainly
due to the fact that the demonstration made by
Zaslavsky\cite{zaslavsky} is incompatible with the existence
  of sporadic randomness.

Here we plan to discuss whether this connection between
$\psi(t)$ and the KS entropy can be extended provided that a form of
non-extensive KS entropy is used to examine the regions of times
shorter than the mean waiting time $T_{av}$.
To discuss this important possibility we must review the recent theoretical
result of Ref.\cite{jin}. This paper provides arguments for the
non-extensive generalization of the Pesin theorem\cite{pesin}.
As shown in Ref.\cite{luigi}, the proper non-extensive
generalization of the Kolmogorov-Sinai entropy
should rest on the calculation of

\begin{equation}
H_{q}(N)\equiv \frac{1-\sum_{\omega_{0}...\omega_{N-1}}
p(\omega_{0}...\omega_{N-1})^{q}}{q-1},
\label{kolmogorov}
\end{equation}
where $p(\omega _{0}...\omega _{N-1})$ is the probability of finding the
cylinder corresponding to the sequence of symbols $\omega _{0}...\omega
_{N-1}$\cite{beck}. In the case where we set $q=1$ the entropy of
Eq.(\ref{kolmogorov})
becomes the ordinary
Kolmogorov-Sinai (KS) entropy. This entropy expression affords
a rigorous way of definining the earlier introduced concept
of \emph{true entropic index Q}. If it
exists, $Q$ is the value of the
entropic index $q$ making the entropy of Eq.(\ref{kolmogorov})
increase linearly in time.

To point out the the importance of the work of Ref.\cite{jin}, we
have to mention the difficulties with the numerical calculation
of Ref.\cite{luigi}. These authors tried to evaluate the dependence
on $N$ of the entropy of Eq.(\ref{kolmogorov}) to establish Q in the 
case of a text
of only two symbols, with strong correlations. They have been forced
to maintain their numerical study within the range of windows
of size $N = 10$.
The case of many more symbols would be beyond the range
of the current generation of computers.
However, when the sequence of symbols is generated by dynamics, as in
the case here under study, and the function $\xi(t,x)$ of Eq.
(\ref{definition}) is available (for convenience, we keep now
explicit the dependence on the initial condition $x$), it is possible
to
adopt the prescription of Ref.\cite{jin}, which writes $H_{q}(t)$ of
Eq.(\ref{kolmogorov}) as

\begin{equation}
H_{q}(t)\equiv \frac{1-\delta ^{q-1}\int dxp(x)^{q}\xi (t,x)^{1-q}}{q-1},
\label{jin}
\end{equation}
where the symbol $t$ denotes time regarded as a continuous variable.
In fact, when
the condition $N>>1$ applies, it is legitimate to identify $N$ with $t$. The
function $p(x)$ denotes the equilibrium distribution density and $\delta $
the size of the partition cells: According to Ref.\cite{jin} the phase
space, a one-dimensional interval, has been divided into $ M = 1/\delta $
cells of
equal size, with $M >>1$.

If the property of Eq.(\ref{superdiffusion}) held true with no
dependence on the initial condition, namely in the form

  \begin{equation}
  \xi(t) = [1 -(z-1)\lambda_{z}t]^{-z/(z-1)},
  \label{superdiffusion2}
  \end{equation}
  one would reach the conclusion that the entropy of Eq.(\ref{jin})
  increase linearly in time if we adopt the entropic index of
  Eq.(\ref{laterdiscussed}). At the same time the theoretical proposal of
  Eq.(\ref{possibility1}) would turn out to be correct. This is the
  reason why we call this special value of $q$ `magic'' and we denote
  it with the capital letter $Q$ thus writing

  \begin{equation}
Q = 1 + (z-1)/z .
\label{firstmagic}
\end{equation}

  What about the case under study, where the local Lyapunov coefficient
  depends on $x$? A possible way out would be given by the following
  procedure. We assign to the entropic index $q$ of Eq.(\ref{jin}) the
  value prescribed by Eq.(\ref{firstmagic}). This has the nice effect of
  making the entropy $H_{q}(t)$ linear dependent on time.
  Unfortunately the distribution function $p(x)$ raised to the power
  of $1 + (z-1)/z$, results in a divergent contribution as it can be
  easily checked noticing\cite{gaspard} that $p(x) \propto 1/x^{z-1}$.
  We notice that it would have been very appealing to establish the
  ``thermodynamic'' foundation of $\psi(t)$ on the relation
  \begin{equation}
  \psi(t) = <(\xi(-t))^{1-Q}>_{Q}^{1/(1-Q)},
  \label{attempt}
  \end{equation}
  with the mean value $<A(x)>_{Q}$ defined by
  \begin{equation}
      <A(x)>_{Q} \equiv \int dx p(x)^{Q} A(x).
      \label{meanvaluedefinition}
      \end{equation}
      This possibility is ruled out by the divergency stemming from
      $p(x)^{Q}$.

      We have to point out another important property suggested by
      Eq.(\ref{jin}). We see that this formula shows that the only
      possible way of making $H_{q}(t)$ independent of the cell size
      $\delta$ is to set $q = 1$. In Ref.\cite{simone} it has been
      pointed out that in the multifractal case it is convenient
      to carry out the average  over the distribution of power indices
      rather than on $p(x)$. This has the nice effect of making
      $H_{q}(t)$ independent of the cell size. In the specific case
      where the invariant distribution exists and is not multifractal,
      as in the case here under study, we are compelled to use $q = 1$.
      This is line with the fact
      that when the invariant distribution of the Manneville map
      exists, namely if $z < 2$, the average Lyapunov coefficient is
      finite, the Pesin theorem work and the ordinary KS entropy can be
      defined\cite{gaspard}.

      Therefore we are left with the possibility of applying
      the prescription of Eq.(\ref{jin}), with the associated average
      of Eq.(\ref{meanvaluedefinition}), to the case where $p(x)$ is
      not the invariant measure, but it is a given out of equilibrium
      distribution. This would correspond to giving up the assumption
      that the Tsallis non-extensive thermodynamics has an equilibrium
      significance, in line with the numerical results of
      Ref.\cite{massi}.

      On the other hand, it is evident, on the basis of the
      conclusions of section III, that the function $\xi(t,x)$ does
      not keep its power-law nature forever, and that in the time scale
      $t >> T_{av}$, with $T_{av}$ given by Eq.(\ref{meanwaitingtime}),
      it is expected to recover an exponential form. Thus, are back to
      the point that this non-extensive form of thermodynamics must be
      temporary.

      We would be tempted to conclude this section with the following
      answer to question (iii) of Section I. The Maxwell's Demon
      thermodynamics is the non-extensive Tsallis thermodynamics, and
      this is a form of out of equilibrium thermodynamics, implying
      that also the Maxwell's Demon effect is temporary.

  This conclusion, to a first sight, seems to be reasonable. However,
  it yields a conflict with the result of the earlier work  of
  Ref.\cite{luigi}. The authors of Ref.\cite{luigi} studied the entropy
  of Eq.(\ref{kolmogorov}) with a symbolic method, and, as earlier
  said, they have been forced to use short-time
  windows. The adoption of short-time windows is imposed by the fact
  they did not make any direct recourse
  to dynamics as in the calculation yielding Eq.(\ref{superdiffusion2}).
The case studied in Ref.\cite{luigi} refers to a dynamic system with 
the same complexity
  as the Manneville map here under study. Probably the main difference
  with the analysis here illustrated is that
  the method adopted by the authors of Ref.\cite{luigi} is equivalent
  to studying the Geisel map rather than the Manneville map, with
  a repartition of the space into only two cells.
  The conflict between the analysis here made and the result of
  the work of Ref.\cite{luigi} is probably due to the fact that here we adopt
  a much finer repartition of the phase space.

This kind of disagreement between two different theoretical predictions
about the entropic index of the same dynamic process is not yet a
reason for the rebuttal of the non-extensive thermodynamics as the
explanation of the Maxwell's Demon effect. In Section V we shall
see that the Maxwell's Demon effect seems to be caused by the
emergence in chamber $A$, made more densely populated by the Demon,
of a diffusion equation with a ``diffusion coefficient''  weaker than
that of the diffusion equation emerging in chamber $B$. This
dynamic effect is explained using the dynamic derivation of L\'{e}vy
statistics, which
implies memory erasure and consequently emerge in a time
scale where the temporary non-extensive thermodynamic regime, if it
ever exists, is over.

\section{L\'{e}vy processes and memory erasure}

Lebowitz\cite{lebowitz} states that the Universe time evolution is
characterized by the transition from a condition of lower to a
condition of larger entropy without implying a departure from
microscopic reversibility.
  Let us focus our attention on the diffusion
  process:
  \begin{equation}
  \frac{dx}{dt} = v(t).
  \label{microscopicdiffusionequation}
  \end{equation}
  Let us assume that the stochastic velocity $v(t)$ fluctuates among
  the two values $W$ and $-W$, and that the distribution of times of
  sojourn in one of these two velocity states, the same for the two
  states, is given by Eq.(\ref{waitingfunction}). It is
  shown\cite{bologna} that the distribution density $p(x,t)$ obeys
  the following equation of motion
  \begin{equation} \label{eq3}
\frac{\partial}{\partial t}p(x,t)= <v^2>
\int_{0}^{t}\Phi_{v}(t')\frac{\partial^2}{\partial x^2}
p(x, t-t')dt'   ,
\end{equation}
where $<v^{2}>$ is the equilibrium mean quadratic value of $v$, and
$\Phi_{v}(t)$ is the normalized correlation function of the variable
$v$. For details, and especially to understand why this equation in
the long-time limit yields the process of  L\'{e}vy diffusion, the
interested reader should consult
Ref.\cite{bologna}.

Here we limit ourselves to noticing a few aspects.
First of all, the distribution density $p(x,t)$ can be
interpreted as a probability concerning a very large number of
particles imbedded in the same space phase, for instance, the phase
space of the
billiards of Zaslavsky\cite{zaslavsky}. Thus, in a sense, the condition
$N\rightarrow \infty$ is fulfilled. This has to do with question (i)
of Section I. If  condition $N \rightarrow \infty$ is
essential for the birth of statistical mechanics, what is the role of
mixing, and that of ensuing ergodicity?

We try to answer this question by
making a second, very important, remark. This is that Eq.(\ref{eq3}) is
exact under
the following conditions:

(a) The variable $v(t)$ is dichotomus, with only two possible values,
either $W$ or $-W$.

(b) All the particles are initially located in the same position,
$x = 0$.

We note that condition (a) is that mentioned in Section IV A
to make easier for the reader to
understand the meaning of the key relation of Eq.(\ref{pi2}), leading
to the entropic derivation of the waiting time distribution $\psi(t)$.
Both conditions are relevant for our main purpose of comparing the
point of view of Zaslavsly\cite{zaslavsky} to that of
Lebowitz\cite{lebowitz}. The latter property certainly realizes an
initial condition whose probability is extremely low, thereby making
impossible the regression to the initial condition. In the ideal case
of a time inversion, this would be possible, in the irrealistic case
of no perturbation of any kind, ranging from the round-off errors to
the environmental fluctuations. This is again in line with the
observations of Lebowitz on the birth of statistical mechanics.

The former property, however, serves in our opinion a purpose which
apparently conflicts with Lebowitz, namely, that of
stressing the importance of
mixing, as a key property leading to an invariant distribution.
For the reader to appreciate this aspect it is necessary that he/she
goes through the derivation of Eq.(\ref{eq3})\cite{allegro}.
The projection method adopted in Ref.\cite{allegro} to derive
Eq.(\ref{eq3}) rests on a projection operator, which, in turns,
implies the existence of an invariant distribution for the
variable $v$. In the specific case of deterministic chaos with mixing,
the prototype of which is given in fact by the Bernouilli shift map
(as we have shown, the Manneville map with z = 1), an invariant
distribution exists and is quickly reached from any off-equilibrium
condition.

The waiting function $\psi(t)$ is proportional to the second derivative
of the correlation function $\Phi_{v}(t)$. This means that
the departure  from the exponential condition
of Eq.(\ref{exponential}) has dramatic effects on the process
of transition to statistical mechanics, as birth of a diffusion
process.
In fact, in the region $1.5 \leq z < 2$ it is still possible to
fulfill the condition necessary for the existence of the invariant
distribution according to the
Kac theorem\cite{KAC}. However, in this specific case
the correlation function $\Phi_{v}$
satisfies the asymptotic property
\begin{equation}
lim_{t \rightarrow \infty} \Phi_{v}(t) = const/t^{\beta},
\label{slowrelaxation}
\end{equation}
with
\begin{equation}
\beta = \frac{z}{z-1} -2.
\label{relation}
\end{equation}
This means that in the interval $1.5 \leq z < 2$ the index $\beta$
runs between $\beta=1$, at the border with the basin of Gaussian
attraction, and $\beta =0$, corresponding to the critical condition
of no relaxation of the velocity correlation function.

In conclusion the condition of
strong chaos, with strongly mixing properties,
has the effect of producing
ordinary Brownian diffusion. It is not clear to us
what would be the consequence of
a dynamic condition with
no form of mixing.
We think that in principle
it is possible to
produce a fast decay of the correlation function
$\Phi_{v}(t)$ as a result of a mere superposition of infinitely many
normal modes (again in action the condition $N \rightarrow \infty$).
However, we are not aware of any treatment proving that
a natural equilibrium distribution is reached also in that case,
if no arbitrary statistical assumptions are made.

We have the impression that in the case of ordinary statistics it
might be
difficult to support the dynamic view against the ($N\rightarrow
\infty$)-perspective, or the latter against the former, without using
very subtle arguments such as the request of neither explicit nor
implicit statistical assumptions, which, in turn, quite probably would
be questioned. For all practical purposes, the two views are expected
to provide equivalent results. For instance, in the dynamic
perspective adopted by Zaslavsly the Poincar\'{e} recurrences are
frequent and those with very short time duration are more probable
than those of very large time duration (see Eq.(\ref{exponential})).
This seems to conflict with Boltzmann's idea that the Poincar\'{e} recurrences
of systems with very large number of freedoms become exceedingly
large, so as to exceed the possibility of any direct observation.
However, if we make the assumption that the experimental observation is
made on a large number of non-interacting particles, all of them
moving in one of the billiards studied by
Zaslavsky\cite{zaslavsky,today,edelman}, we
reach a different conclusion. The phase space under study is not more
that of the two-dimensional billiard: It becomes a 2N-dimension phase
space. The distribution density driven by Eq.(\ref{eq3}) becomes
a $2N$-dimensional trajectory, whose return to the initial condition
takes place, in accordance with Boltzmann's view, virtually after an
infinitely long time. In Zaslavsky's picture the process
corresponding to the $2N$-dimension trajectory leaving the initial
condition and returning to it after an extremely long time would
correspond to the following scenario. Let us imagine, for instance,
that the initial distribution
density is different from zero only in a small fraction of the
whole $2$-dimension phase space, where it is assumed to be constant.
As a result of mixing, this initial distribution, throughout its time
evolution,  would undergo
a fragmentation process that, within a kind of coarse-grained
perspective, would correspond to the volume of this initial condition
growing till to becoming identical to the volume of the whole phase
space. The Poincar\'{e} recurrences of the Boltzmann-Lebowitz
perspective would correspond to this dilated distribution shrinking to
the initial condition, a process that is beyond the range of the
experimental observation as the return of the Boltzmann-Lebowitz
trajectory.

We think that the relevance of dynamics for statistical mechanics
cannot be easily ruled out in the case of processes with long-time
memory as those here under discussion. Quite on the contrary, here
we show  that the dynamic approach to L\'{e}vy diffusion in
the form discussed in Ref.\cite{bologna} can account for the
Maxwell's Demon effect. In this moment, it is not clear to us how this
interesting effect might be explained using the $(N\rightarrow
\infty)$-perspective. The authors of Ref.\cite{bologna}  prove that in
a time scale much larger than $T_{av}$ Eq.(\ref{eq3}) as an effect of
the memory erasure provoked by sporadic randomness becomes equivalent
to an equation of motion that in the Fourier representation
reads
\begin{equation}
     \frac{\partial}{\partial t} \hat{p}(k,t) = - b |k|^{\beta+1}\hat{p}(k,t),
     \label{memoryerasureeffect}
     \end{equation}
     with
     \begin{equation}
	b \equiv
	\Gamma(-\beta) \, W^{\beta+1} \, T_{av}^{\beta} \,\beta^{(\beta+1)} \, cos(\frac{\beta}{2} \pi).
	\label{diffusioncoefficient}
	\end{equation}
	Note that $\hat{p}(k,t)$ is the Fourier transform of $p(x,t)$ and
	that this is the Fourier representation of the well known process of
	L\'{e}vy diffusion. The element of interest of the result of the
	dynamic approach of Ref.\cite{bologna} is that the coefficient
	$b$ of Eq.(\ref{diffusioncoefficient}) plays the role of a diffusion
	coefficient. The reader can easily realize that the intensity of
	this coefficient can be different in the two chambers of the
	experiment on the Maxwell's Demon effect\cite{zaslavsky,today,edelman}
	even if the velocity $W$ is the same. The possibility of realizing
	two different values for $b$ using the same kinetic energy becomes
	still wider if we refer to a generator of fluctuations with an
	inverse power law distribution of waiting times more general than the
	Manneville map\cite{note}. It is evident that two different diffusion
	equations in the two chambers can produce a breakdown of the
	condition of equal population. The trajectory, in fact, will tend to
	spend more time in the regions with a smaller ``diffusion
	coefficient''.

In conclusion, according to this heuristic interpretation, the
Maxwell's Demon Effect is permanent and it can show up after the long
time process of transition from dynamics to diffusion. This rules out
the non-extensive thermodynamics of Tsallis as a possible theoretical
interpretation of this effect. In fact, in Section IV B we have seen
that if the non-extensive approach applies, it does as a form of
temporary thermodynamics.

We can also rule out the non-extensive nature of entropy as the basic
property behind the Maxwell's Demon Effect. Let us see why.
We have two chambers, $A$ and $B$, separated
by a wall, with a little hole.
$A$ can be a Sinai billiard and $B$ a Cassini billiard\cite{edelman}.
The physical entropy of this regime, $S$, expressed in terms of the 
conventional
Gibbs prescription, might not be extensive. The entropic indicator
advocated by Tsallis\cite{brazil} is not additive, and it violates the property

\begin{equation}
S(A+B) = S(A) + S(B)
\label{additivity}
\end{equation}
even when the two subsystems $A$ and $B$ are totally uncorrelated
the one from the other. In the case of the dynamic approach to L\'{e}vy
diffusion the breakdown of the additivity condition might be dictated
by memory effects. To explain why it is so with intuitive arguments,
we can refer ourselves to the dynamical experiment itself used by
Zaslavsly and Edelman\cite{edelman}  to reveal the existence of
Maxwell's Demon effect.
A particle that moves from $A$ into $B$ might still have memory
of its initial condition in $A$, if  sporadic randomness
did not act for a sufficiently extended time.
This means that the transition process that
in the long-time scale will produce
L\'{e}vy statistics is not extensive in nature.
However, we expect that when L\'{e}vy statistics are finally established,
the usual additive condition is restored, as a consequence
of the  memory erasure process. Using the earlier arguments we can
imagine the possibility that the two chambers might correspond to two
different difusion equations, even if the velocity intensity of the particle
does not change with moving from the one to the other chamber, thereby
resulting in the Maxwell's Demon effect after the process of memory
erasure.

\section{concluding remarks}
Let us summarize our conclusion about the questions raised
in Section I.

Question (i).
Our arguments concerning question (i) are not compelling and
rigorous. However, we hope that they might make plausible
the following statement:
Both deterministic chaos, whose role is well understood
only in the  case of low-dimension systems, and
the condition $N \rightarrow \infty$
give rise to the birth of statistical mechanics.
We are convinced that the joint use of these
two perspectives
will contribute a deeper understanding
of the dynamic origin of statistical mechanics.
It is interesting to notice that the exponential
of Eq.(\ref{exponential}), implying a frequent return to the initial
condition, does not conflict with the irreversible character of
thermodynamics and statistical mechanics if we perceive
the Gibbs ensemble method as being equivalent
to the $(N \rightarrow \infty)$-perspective.

Question(ii). We found the proposal of Eq.(\ref{possibility2}),
valid at the exit time of a trajectory from the laminar region,  to be
the correct connection between the sensitivity to initial
conditions and the distribution of waiting times in the laminr
region. This means that in the case of the dynamical systems studied
by Zaslavsky, the distribution of the Poincar\'{e} recurrence times
can be related to the KS entropy only in the absence of intermittency.
When an intermittent process is present, the long-time form
of the distribution $P_{R}(t)$
changes from the exponential form of  Eq.(\ref{exponential}) to an
inverse power law form. In the region $1.5 \leq z \leq 2$,
corresponding to the emergence of L\'{e}vy processes, the KS entropy
is still finite and the adoption of Eq.(\ref{exponential}) would conflict
with the observed inverse power law nature of this distribution.
The non-extensive thermodynamics of Tsallis\cite{brazil} might
help, provided that it is viewed as temporary and provided that the conflict
between the prediction about whether $Q>1$ or $Q<1$ is settled.

Question (iii). The dynamic approach to L\'{e}vy diffusion of
Ref.\cite{bologna} proves that the Maxwell's Demon effect is
permanent and it takes place in a long-time scale. This means that it
is not necessary to use memory properties to explain it, and
consequently, not even the non-extensive nature of entropy.

Question (iv). The condition of ordinary statistical mechanics does
not imply an abrupt transition when we set $z = 1$. However, for even
infinitesimally small deviations from $z = 1$ in the long time limit
the waiting time distribution $\psi(t)$ appears to be characterized
by an inverse power law form. This does not have strong consequences,
though, because in the region $z < 1.5$ the second moment of the
waiting time distribution is finite, and consequently the system is
attracted by the Gauss basin of attraction. Nevertheless, the waiting
time distribution in the region where $z$ is very close to $1$, is an
exponential at short time and an inverse power law at long time, a
property which makes it incompatible with the structure generated
by the method of entropy maximization resting on the non-extensive
indicator of Tsallis\cite{brazil}.

It seems to us that the long-standing problem of a thermodynamic approach
to L\'{e}vy processes\cite{shlesinger} is not yet solved, and that
the Maxwell's Demon effect of Refs.\cite{today,zaslavsky,edelman}
can be regarded as a paradigmatic case challenging the Boltzmann
perspective advocated by Lebowitz\cite{lebowitz}. We plan to do
further research work along these directions.


\begin{references}
\bibitem{note} We note that the work of Ref.\cite{luigi}
is based on randomly drawing a number $x$ in the interval $[0,1]$,
with the Bernouilli map. With
a nonlinear transformation this number is changed into a number
$y$ with an inverse power law distribution. Then they toss a coin
and decide whether to adopt the symbol $+$ or the symbol $-$.
They consider the integer
part of $y$, $[y]$, and fill [y] sites of an infinite seuence
with either the symbol $+$ or
the symbol $-$, according to the indication of the tossed coin.
The work of Ref.\cite{marco} is devoted to the study of the
complexity of the resulting sequence.
The process of coin tossing and number drawing is carried out at
regular time intervals. This is equivalent to using a time $N$,
corresponding to the number of times the process of coin tossing and
number drawing has been done. The entropy increase is $S(N)
= 2 N log 2$. In fact, $log 2$ is the uncertainty related to both coin
tossing and to a time step of the Bernouilli map.
If the sequences of $+'s$ and $-'$ is fixed without tossing a coin
and alternating the $+'s$ with $-'s$, then the entropy increase is
$S(N) = N log 2$. The complexity of the Manneville map depends on the
connection between the ``physical'' time $Y(N) = [y]_{1} + [y]_{2}\ldots
[y]_{N}$ and the ``internal'' time $N$. The connection between
these two times is the essential property determining the Kolmogorov
complexity of both kinds of maps, the difference bing only a factor
of 2. The essential aspect  of the
complexity of the Manneville map studied by Gaspard and
Wang\cite{gaspard} is the transition from a non vanishing
KS entropy with $z>2$ to a vanishing KS entropy with $z<2$. For the
meaning of $z$ see Eq.(\ref{manneville}).
\end{references}
\end{document}